\documentclass[11pt]{article}
\usepackage{amssymb}
\usepackage{amsmath}
\usepackage[dvips]{epsfig}

\def\##1{\underline #1}
\def\=#1{\underline{\underline #1}}

\def\eps{\epsilon}
\def\epso{\epsilon_0}
\def\muo{\mu_0}
\def\ko{k_0}
\def\lambdao{\lambda_0}
\def\etao{\eta_0}
\def\.{\mbox{ \tiny{$^\bullet$} }}

\def\epsr{\epsilon_r}
\def\mur{\mu_r}
\def\etar{\eta_r}
\def\tz{{\tilde\zeta}}

\def\ux{\#{u}_x}
\def\uy{\#{u}_y}
\def\uz{\#{u}_z}

\def\le{\left(}
\def\ri{\right)}
\def\les{\left[}
\def\ris{\right]}
\def\lec{\left\{}
\def\ric{\right\}}

\def\c#1{\cite{#1}}

\def\r#1{(\ref{#1})}

\begin{document}

\noindent {\em Spectral response of 
 Cantor multilayers made of
materials with negative refractive index}

\bigskip

\noindent Jaline Gerardin {\em and}  Akhlesh Lakhtakia\footnote{Corresponding
author. Tel: +1 814 863 4319; Fax: +1 814 865 9974; E--mail: AXL4@PSU.EDU} 

\bigskip

\noindent  CATMAS~---~Computational and Theoretical Materials Science
Group\\ Department of Engineering Science and Mechanics\\
Pennsylvania State University, University Park, PA 16802--6812, USA
\bigskip

\bigskip

\noindent ABSTRACT--Whereas Cantor multilayers made of an isotropic
dielectric--magnetic material with positive refractive
index will show power--law characteristics,
low--order Cantor multilayers made of materials with negative refractive
index will not exhibit the power--law nature. 
A reason for this anomalous behavior
is presented.



\section{Introduction}\label{S:intro}
This letter addresses the incorporation of isotropic
materials with negative refractive index  \c{SSS}, \c{MLW}
 in fractal filters
inspired by Cantor dusts \c{Man}, \c{LMVV}.

The emergence of Cantor dusts, bars and cakes during the late 19th
century has been described at some length by Mandelbrot \c{Man}.
Briefly, the simplest Cantor dust is formed by dividing
the closed interval $\les 0,\,1\ris$ into 3 pieces and 
removing the center open piece $\le  1/3,\,2/3 \ri$,
 repeating the trifurcation--and--removal
process on the remaining intervals $\les 0,\,1/3\ris$
and $\les 2/3,\,1\ris$, and continuing
in that fashion {\em ad infinitum\/}. The fractal
(similarity) dimension of the resulting dust is $\log 2/\log 3\sim
0.6309$. Similar structures in $p$--dimensional space, ($p=1,\,2,\,...$),
can be constructed {\em via\/} spatial convolution \c{LMVV}. In
particular, the constructs called Cantor bars appear to have
captured the imagination of optical--filter researchers, as recounted
recently by Lehman \c{Leh2002}, because of their
putatively self--similar response properties in the frequency domain
\c{LZSG}.

The materials of choice for optical Cantor filters are
isotropic dielectric with relative permittivity $\epsr$.
Although $\epsr$ is a complex--valued function of frequency,
the usual practice in optics is
to ignore dissipation by setting ${\rm Im}\les \epsr\ris = 0$.
In the area
of fractal optics, with emphasis still on understanding basic
interactions in nonperiodic multilayers, dispersion is also
ignored \c{Leh2002}--\c{BMS}. The structural self--similarity
of the Cantor bars is then expected to result in the self--similarity
of the spectral reflectance/transmittance responses
of optical Cantor filters
to normally incident light \c{LZSG}. Truly, physically realizable
Cantor filters are not actually fractal but 
pre--fractal instead \c{LC}~---~
so that the spectral self--similarity can only be approximate
\c{LehG}.

On examining the available literature, two questions
arise. First,
will the situation change for Cantor filters made of
isotropic dielectric--magnetic materials (with relative
permeability denoted by $\mur>1$)? Second, will the situation
change if both $\epsr < 0$ and $\mur < 0$?

The second question arose because of the supposed verification
of the existence of  negative refractive index (NRI)  by
Shelby {\em et al.\/} \c{SSS} last year. Experiments
performed on certain composite
materials with oriented microstructure
suggested that these materials
are endowed with  negligible dissipation as well as NRI
 in some appreciably wide frequency band in the
centimeter--wave regime. Also called {\em left--handed materials\/}
by some researchers (despite possessing no handedness), in  NRI
materials the phase velocity  is pointed opposite to the
direction of energy flow (and attenuation) \c{MLW}, \c{Rup}.
Although the extant experimental results
 are not perfect \c{Lakh0}, \c{NGV}, the
essential conclusion of the existence of NRIs
 appears undeniable. As NRIs can
potentially lead to exciting new technologies \c{Pen}, theoretical
consideration is warranted.

In this letter, we answer the two questions posed earlier
in a unified way. Section 2 is devoted to the theory of reflection
and transmission of normally incident plane waves by
Cantor multilayers. Numerical results are presented
and discussed in Section 3.

\section{Theory}

A  Cantor multilayer is constructed sequentially
as follows: Take a layer of thickness $\ell_0$
made of a certain material with $\epsr$ and
$\mur$ as its constitutive parameters. Call this layer a multilayer
of order $N=0$. Next, cascade two multilayers of order $N=0$
 inserting a  space of thickness $\ell_0/f$,
$f \geq 1$, in between. Call this a multilayer of order $N=1$.
Its total thickness $\ell_1 = (2+1/f)\ell_0$. Continue
in this manner. Thus, a multilayer of order $N+1$ is formed
by inserting a space of thickness $\ell_{N}/f$
between two multilayers of order $N$. The thickness of a
multilayer of order $N+1$ is then $\ell_{N+1} = (2+1/f)\ell_{N}
=(2+1/f)^{N+1}\ell_0$. The fractal dimension of the multilayer
is  given by
\begin{equation}
{\cal D} =\frac{ \log 2}{\log(2+1/f)}\,,
\end{equation}
which concept is applicable strictly  in the limit $N\to\infty$.

Let a Cantor multilayer of order $N$ occupy the space
$0\leq z\leq \ell_N$.
Suppose a plane wave is normally incident on this multilayer from the
vacuous half--space
$z \leq 0$, with $\lambdao$ denoting its
wavelength. Therefore, a reflected plane wave also exists
in the same half--space. Furthermore,
a transmitted plane wave is engendered
in the vacuous half--space $z\geq \ell_N$. The corresponding electric field
phasors are given by
\begin{equation}
{\#E}(z) = \ux\,
\lec
\begin{array}{ll}
 \exp(i\ko z) + \rho_N\exp(-i\ko z)\,,&\qquad z \leq 0
\\
\tau_N \exp \les i\ko (z-\ell_N)\ris\,,&\qquad z \geq \ell_N
\end{array}\right.,
\end{equation}
where $\ko=2\pi/\lambdao$ is the 
wavenumber in vacuum;  $\rho_N$ and $\tau_N$ are
the reflection and the transmission coefficients, respectively,
both complex--valued;
and $(\ux,\uy,\uz)$ is the triad of cartesian unit vectors.
An $\exp(-i\omega t)$ time--dependence is implicit, where $\omega
=\ko /(\epso\muo)^{1/2}$
is the angular frequency, while $\epso$
and $\muo$ are the permittivity and the permeability
of vacuum, respectively.

The coefficients $\rho_N$ and $\tau_N$ can be 
easily determined using
a 2$\times$2 matrix algebra
\c{Lopt}. After defining the two matrixes
\begin{equation}
{\sf A} = \les \begin{array}{cc} 0 &\muo  \\ \epso  & 0
\end{array}\ris\,,\qquad
{\sf B} = \les \begin{array}{cc} 0 &\muo\mur \\ \epso\epsr & 0
\end{array}\ris\,,
\end{equation}
the matrixes ${\sf M}_p$, $0\leq p\leq N$, are
iteratively computed as
\begin{equation}
{\sf M}_{p+1} = {\sf M}_{p}\. e^{i\omega (\ell_p/f){\sf A}}\.
{\sf M}_{p}\,,\quad 0\leq p\leq N-1\,,
\end{equation}
beginning with
\begin{equation}
{\sf M}_0 = e^{i\omega \ell_0{\sf B}}\,.
\end{equation}
The boundary value problem for the electromagnetic fields
then involves the solution of the equation
\begin{equation}
\tau_N \, \les\begin{array}{c} 1 \\ 
\etao^{-1} \end{array}\ris={\sf M}_N\.
\les\begin{array}{c} (1+\rho_N) \\\etao^{-1} (1-\rho_N) \end{array}\ris\,,
\end{equation}
where $\etao =  (\muo/\epso)^{1/2}$ is the intrinsic
impedance of vacuum.
The principle
of conservation of energy entails that $\vert\rho_N\vert^2
+\vert\tau_N\vert^2 \leq 1$, with the equality coming in
when the multilayer is made of a non--dissipative
material.

\section{Numerical results and discussion}
Following normal practice, we implemented the
foregoing equations to compute $\rho_N$ and $\tau_N$
for non--dissipative and non--dispersive materials. 
We varied the quantity $\zeta =\ko\ell_0$ for various
values of $N$, while keeping $\epsr$ and $\mur$
fixed. 

The spectrums of $\vert\rho_N\vert^2$ and $\vert\tau_N\vert^2$ turned
to be identical to the ones reported in the literature \c{Leh2002},
\c{LZSG}
for optical Cantor filters (i.e., with  $\epsr>1$
and $\mur=1$).
Those for $\lec \epsr>1,\,\mur>1\ric$
and $\lec\epsr<0,\,\mur <0\ric$ turned to be
qualitatively similar,
and therefore do not need reproduction here.

As $\zeta$ increases from zero, the fundamental layer
thickness $\ell_0$ becomes an increasingly significant
fraction of the wavelength $\lambdao=2\pi/\ko$, and eventually
surpasses $\lambdao$. In other words, layers are
electrically thin for small $\zeta$, and an increase in
$\zeta$ amounts to magnification. Therefore we
evaluated the value $\tz_N$ of $\zeta$ at which the first
minimum of $\vert\tau_N\vert$ occurs as $\zeta$ increases
from zero, thereby reckoning $\tz_N$ as
a reasonable parameter containing structural information
on the chosen multilayers. If indeed 
the structural self--similarity
of  Cantor multilayers would result in their spectral self--similarity,
we expect the
relationship
\begin{equation}
\label{fr}
\tz_N = 2^{-N/{\cal D}}\,\tz_0 
\end{equation}
to emerge from our numerical investigations.

Figures 1 and 2 contain plots of $\log\tz_N$
versus $N$ for Cantor multilayers
made with positive refractive index (PRI) materials
($\lec\epsr=3,\,\mur=1.02\ric$
or $\lec\epsr=4,\,\mur=1.02\ric$), and for
Cantor multilayers made with their
NRI analogs
($\lec\epsr=-3,\,\mur=-1.02\ric$
or $\lec\epsr=-4,\,\right.$ $\left.\mur=-1.02\ric$). 
The factor $f=1$ for Figure 1, and $f=2$ for Figure 2.
Two conclusions
can be immediately drawn
from these two figures as follows:
\begin{itemize}
\item[A.] The relationship $\tz_N = 2^{-N/{\cal D}_{PRI}}\,\tz_0$
satisfied by Cantor multilayers with  PRI materials
is  a power law with  ${\cal D}_{PRI}> {\cal D}$,
and could be fractalesque \c{Lsst,ABLM}.
\item[B.] The data for Cantor multilayers with NRI
materials indicates two different regimes, one for small $N$
and the other for large $N$, the second regime characterized by
a power law.
\end{itemize}
The foregoing conclusions suggest that the effect of NRI
materials on electromagnetic
fields must be substantively different from that of PRI materials,
for the anomalous first regime to arise for Cantor multilayers
with NRI materials.
Furthermore, in the present context,
 the difference must be evident definitely for order
$N=0$. 

Hence, we analyzed the planewave response
of a single layer to obtain
\begin{equation}
\label{rho0}
\rho_0= \frac{(\etar^2-1)\sin\beta}{(\etar^2+1)\sin\beta
+2i\etar\cos\beta}\,
\end{equation}
and
\begin{equation}
\label{tau0}
\tau_0= \frac{2i\etar}{(\etar^2+1)\sin\beta
+2i\etar\cos\beta}\,.
\end{equation}
Here, the relative impedance $\etar=+\sqrt{\mur/\epsr}$
must  be positive real,  while
the sign 
of $\beta=\ko\ell_0\sqrt{\mur\epsr}$ has to be
positive/negative for PRI/NRI materials \c{MLW,JL}.
Denoting the phase of a complex number $\xi$ by $\angle \xi$,
we conclude from the
foregoing equations that
\begin{equation}
\label{ch}
\lec \epsr\to-\epsr\,,\mur\to-\mur\ric \,\Rightarrow\,
\lec \vert\rho_0\vert\to  \vert\rho_0\vert,\,
\vert\tau_0\vert\to  \vert\tau_0\vert,\,
 \angle \rho_0\to-\angle \rho_0\,,
 \angle \tau_0\to-\angle \tau_0\ric\,.
\end{equation}

In light of the relationship \r{ch}, let us compare a PRI layer
and a NRI layer~---~labeled $a$ and $b$, respectively~---~
such that 
$\mu_{r_a}=-\mu_{r_b}
 > 0$ and $\eps_{r_a}=-\eps_{r_b}>0$, while
the wavenumber $\ko$ is fixed.
If the thicknesses of the two layers are such
that the sum $\beta_a+\vert\beta_b\vert$ is an integral
multiple of $2\pi$, then \r{rho0} and \r{tau0}
yield
$\rho_{0_a} =  \rho_{0_b} $ and
$\tau_{0_a}=\tau_{0_b}$. Thus,
a PRI layer of
a certain thickness is {\em equivalent\/}
 to a NRI layer of different thickness,
in terms of the complex--valued reflection and transmission
coefficients at a fixed wavelength.
But the thickness of the equivalent NRI layer is
wavelength--dependent~---~which implies that
{\em a PRI Cantor multilayer
 is equivalent at different
wavelengths to different NRI Cantor
multilayers.\/}
Not surprisingly therefore, the spectral characteristics
of a PRI and a NRI Cantor multilayers with the same
$\ell_0$ are not isomorphic.

The difference is very noticeable for small $N$ in Figures 1
and 2. As $N$
increases, the value of $\tz_N$ decreases for both PRI
and NRI multilayers~---~in other words,  the fundamental layer
of thickness $\ell_0$ becomes
electrically thinner at the first transmittance
minimum and,  therefore, a weaker reflector as
well as a stronger transmitter.
Structural characteristics then dominate
over the consequences of \r{ch},  because 
$\vert\rho_0\vert\simeq 0$ and $\vert\tau_0\vert\simeq 1$.
As the difference between PRI and NRI multilayers
lessens with increasing $N$, the latter also begin
to evince power--law characteristics.

The crossover between the anomalous and the
power--law regimes for NRI
multilayers takes place at a higher value of $N$
as $f$ increases. This general trend
is indicated by Figures 1 and 2 as well as calculations
for other values of $f$.

To conclude, we have shown that the planewave reflection and
transmission spectrums of a Cantor multilayer made of an isotropic
dielectric--magnetic material with positive refractive
index shows power--law characteristics 
which indicate spectral self--similarity. However,
if the same multilayer were to be made of a material
with negative refractive index, then the power--law nature
is not going to be evident when the interaction
between the material layers and the interleaving vacuous spaces
is substantial (the small--$N$ regime). The existence of this
anomalous regime can be attributed to the 
reflection/transmission phase reversal of a NRI layer
in relation to its PRI analog. If that interaction
is insubstantial (the large--$N$ regime), the structural
features would dominate the constitutive features, and
the power--law characteristics would be evident also for
the NRI Cantor multilayer. 

\noindent {\bf Acknowledgement} We thank an anonymous
reviewer for the suggestion to examine the large--$N$
responses.

\begin{figure}[!ht]
\centering \psfull \epsfig{file=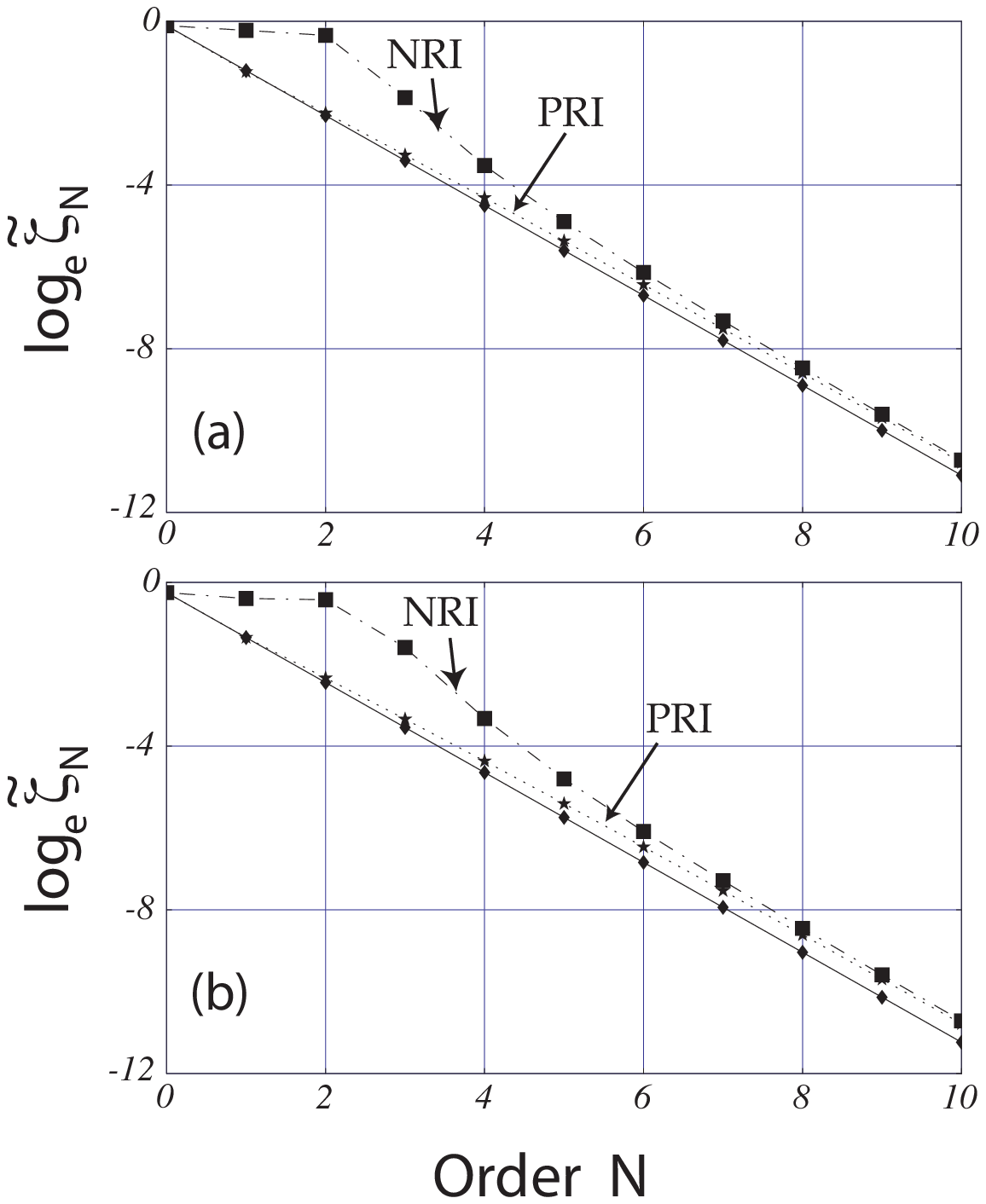,width=4in}
\caption {Calculated
values of $\log\tz_N$ for Cantor multilayers of
orders $N$ when $f=1$. 
Dotted lines join points  for PRI Cantor multilayers, dashed-dotted
lines  for NRI Cantor multilayers,
and solid lines  for $\tz_N=2^{-N/{\cal D}}\tz_0$.
(a) $\epsr=\pm 3$ and $\mur=\pm 1.02$;
(b) $\epsr=\pm 4$ and $\mur=\pm 1.02$.
}
\end{figure}

\begin{figure}[!ht]
\centering \psfull \epsfig{file=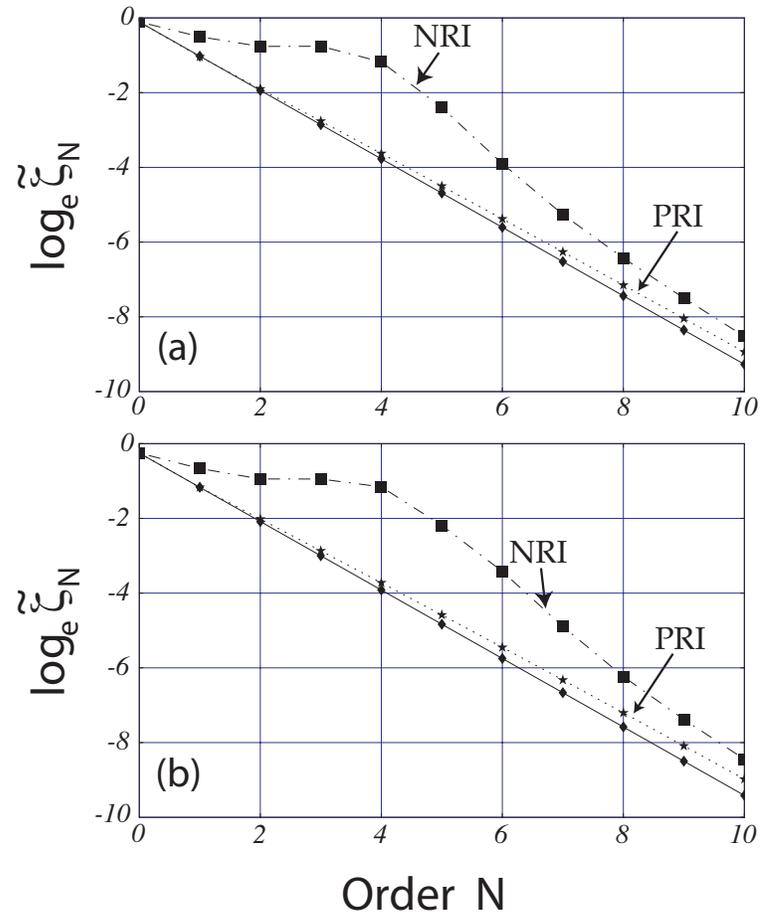,width=4in}
\caption {Same as Figure 1, but for $f=2$.}
\end{figure}

\end{document}